\documentclass[aps,prd,twocolumn,showpacs,amsmath,amssymb]{revtex4-1}
\usepackage{graphicx}
\usepackage{subfig}
\usepackage{caption}
\usepackage{epstopdf}
\usepackage{color}
\usepackage{multirow}
\usepackage{setspace}
\usepackage{overpic}
\usepackage{amsmath}
\usepackage{amssymb}
\usepackage[bookmarksnumbered, pdfstartview=FitH,colorlinks,urlcolor=blue, citecolor=blue,linkcolor=blue] {hyperref}
\usepackage{lineno}
\usepackage{bm}
\usepackage{rotating}
\usepackage{verbatim}
\usepackage[utf8]{inputenc}
\graphicspath{{Fig/}}
\usepackage{bigstrut}
\begin{document}

%\linenumbers

\title{\boldmath Search for the lepton number violating process $J/\psi \to K^+K^+e^-e^- +c.c.$}

\author{%% Saved at => 2025-02-15
\begin{small}
\begin{center}
M.~Ablikim$^{1}$, M.~N.~Achasov$^{4,c}$, P.~Adlarson$^{77}$, X.~C.~Ai$^{82}$, R.~Aliberti$^{36}$, A.~Amoroso$^{76A,76C}$, Q.~An$^{73,59,a}$, Y.~Bai$^{58}$, O.~Bakina$^{37}$, Y.~Ban$^{47,h}$
% ...
\\
\vspace{0.2cm}
(BESIII Collaboration)\\
\vspace{0.2cm} {\it
$^{1}$ Institute of High Energy Physics, Beijing 100049, People's Republic of China\\
% ...
$^{4}$ Budker Institute of Nuclear Physics SB RAS (BINP), Novosibirsk 630090, Russia\\
% ...
$^{36}$ Johannes Gutenberg University of Mainz, Johann-Joachim-Becher-Weg 45, D-55099 Mainz, Germany\\
$^{37}$ Joint Institute for Nuclear Research, 141980 Dubna, Moscow region, Russia\\
% ...
$^{47}$ Peking University, Beijing 100871, People's Republic of China\\
% ...
$^{58}$ Southeast University, Nanjing 211100, People's Republic of China\\
$^{59}$ State Key Laboratory of Particle Detection and Electronics, Beijing 100049, Hefei 230026, People's Republic of China\\
% ...
$^{73}$ University of Science and Technology of China, Hefei 230026, People's Republic of China\\
% ...
$^{76}$ University of Turin and INFN, (A)University of Turin, I-10125, Turin, Italy; (B)University of Eastern Piedmont, I-15121, Alessandria, Italy; (C)INFN, I-10125, Turin, Italy\\
$^{77}$ Uppsala University, Box 516, SE-75120 Uppsala, Sweden\\
% ...
$^{82}$ Zhengzhou University, Zhengzhou 450001, People's Republic of China\\
\vspace{0.2cm}
$^{a}$ Deceased\\
% ...
$^{c}$ Also at the Novosibirsk State University, Novosibirsk, 630090, Russia\\
% ...
$^{h}$ Also at State Key Laboratory of Nuclear Physics and Technology, Peking University, Beijing 100871, People's Republic of China\\
% ...
}
\end{center}
\vspace{0.4cm}
\end{small}
}

\begin{abstract}
Based on $(10087\pm 44)\times10^{6}$ $J/\psi$ events collected with the BESIII detector, we search for the lepton number violating decay $J/\psi \to K^+K^+e^-e^- +  c.c.$  for the first time. The upper limit on the branching fraction of this decay is set to be  $2.1 \times 10^{-9}$ at the 90$\%$ confidence level with a frequentist method. This is the first search for $J/\psi$ decays with the lepton number change by two, offering valuable insights into the underlying physical processes.
\end{abstract}

\maketitle

%----------------------------------------------------
\section{Introduction}
%----------------------------------------------------
The absence of a right-handed neutrino component in the Standard Model (SM),  coupled with the imperatives of $SU(2)_L$ gauge invariance and renormalizability, results in the prediction that neutrinos should be massless. However, experimental evidence from neutrino oscillation studies~\cite{sm1,sm2,sm3,sm4} compellingly demonstrates that neutrinos possess a non-zero, albeit minute, mass. This discrepancy hints the existence of physics that extends beyond the current understanding provided by the SM.

In 1937, Majorana proposed that neutrinos might be their own antiparticles, a scenario in which the neutrino and the antineutrino are indistinguishable~\cite{majorana}. This possibility raises an unresolved question: are neutrinos Dirac particles, distinct from their antiparticles, or Majorana particles, which are not?  The ``see-saw" mechanism~\cite{see1,see2,see3} is a popular model for generating light neutrino masses, requiring a heavy Majorana neutrino with a mass ranging from a few hundred MeV/$c^2$ to a few GeV/$c^2$. If neutrino masses arise from this mechanism, there could be processes that violate lepton number by two units ($\Delta L=2$), revealing the Majorana nature of neutrinos. Thus, search for lepton number violation (LNV) is crucial to confirm whether neutrinos are Majorana particles. In 1939, based on Majorana's theory, Furry~\cite{furry} studied the neutrinoless double beta ($0\nu2\beta$) decay~\cite{beta}, which is an LNV process. Since then, various theories concerning LNV processes have been proposed, and numerous experiments have been carried out to explore these processes.

In addition, the observed matter-antimatter asymmetry in the Universe presents a significant challenge to our understanding of nature. The Big Bang theory, which is the prevailing cosmological model for the Universe's evolution, predicts equal amounts of baryons and anti-baryons in the early Universe. However, the observed amount of baryons exceeds that of the anti-baryons  by a factor currently estimated at $10^9 - 10^{10}$~\cite{bn}. To give a reasonable interpretation of the baryon-antibaryon asymmetry, Sakharov proposed three principles~\cite{sa}, the first of which is that baryon number conservation must be violated. Many theories believe that if there is baryon number violation, there will also be LNV\cite{BL-ther}.  Consequently, experimental searches for the LNV $\Delta L = 2$ processes are of great interest.

Apart from the $0\nu2\beta$ decay experiments, which could search for Majorana neutrinos, several collider experiments have explored related processes. For example, the LHCb experiment investigated the $B^{-} \to \pi^{+} \mu^{-} \mu^{-}$ decays~\cite{LHCb}, the CMS experiment studied the $p\bar{p}\to e^+e^+ X$ process~\cite{CMS}, the BaBar experiment analyzed the $D^{+}_{s} \to \pi^{+} l^{+} l^{+}$ decays~\cite{BaBar}, and the ATLAS experiment searched for right-handed $W$ bosons and heavy right-handed Majorana or Dirac neutrinos, using final state containing a pair of charged leptons and two jets~\cite{Atlas}. The Gerda experiment searched for the rare $0\nu2\beta$ decay process in $^{76}Ge$~\cite{Gerda}. In addition, the CLEO experiment investigated the $D \to \pi(K) e^{+}e^{+}$ decays~\cite{Cleo} and the FOCUS experiment searched for the decays of $D^{+}$ and $D^{+}_{s}$ containing the same-sign di-muons~\cite{focus}. In recent years, the BESIII experiment searched for the LNV decays of charmed mesons, including $D \to K \pi e^{+} e^{+}$~\cite{LNV1}, $D^{0} \to \Bar{p}e^{+}(pe^{-})$~\cite{LNV2}, $D^{+} \to \Bar{\Lambda}(\Bar{\Sigma}^{0}e^{+}) $~\cite{LNV3}, and $D^+(D^-)\to ne^+(\bar n e^-)$~\cite{LNV4}. In all above searches, no evidence has been found.

\begin{figure}[ht]
\centering
\includegraphics[scale=0.14]{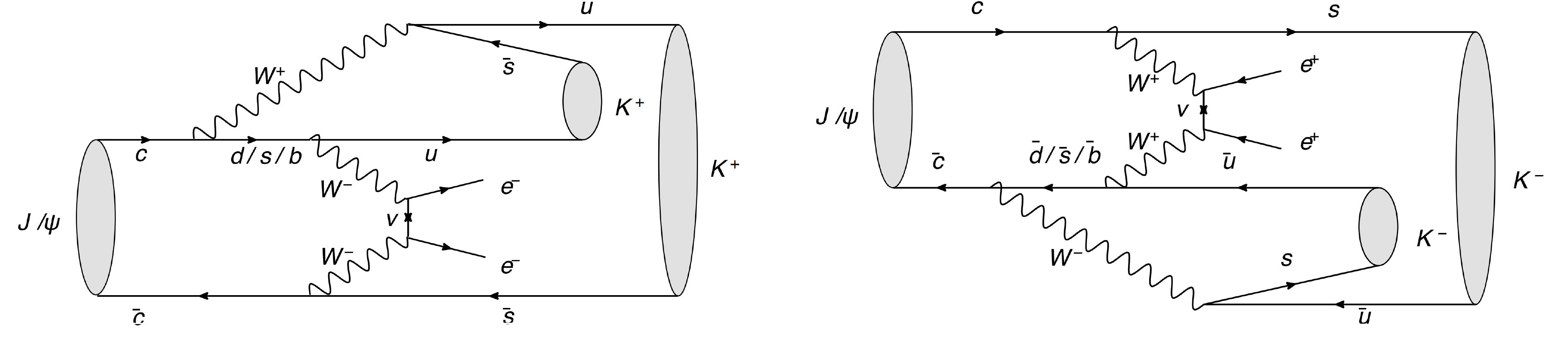}
\caption{Possible Feynman diagrams of $J/\psi \to K^{+}K^{+}e^{-}e^{-}+c.c$.}
\vspace{0.05\baselineskip}
\label{fig::feynman}
\end{figure}

In the $\tau$-charm energy region, the huge data set of charmonia (such as $J/\psi$) can be analyzed to search for various LNV processes, and it is feasible to search for $J/\psi \to K^+K^+e^-e^-$ (throughout this paper, charge conjugation is implied) and other LNV processes in the decay products of $\phi, \omega, K^0, \eta/\eta'$, and so on. Fig.~\ref{fig::feynman} shows possible Feynman diagrams of $J/\psi \to K^+K^+e^-e^-+c.c$. Search for this process would provide hints for new physics, as they are expected to be suppressed due to higher-order processes in the SM.

In this paper, by analyzing $(10087\pm 44)\times10^{6}$ $J/\psi$ ~\cite{ref::jpsi_num}~events collected with the BESIII  detector~\cite{BESIII_intro} operating at the BEPCII storage ring, we search for the LNV decay $J/\psi \to K^+K^+e^-e^-$ for the first time. Furthermore, we explore the upper limits on the branching fractions of this decay under different Majorana neutrino mass hypotheses. To avoid possible bias, a blind analysis technique is employed, where the data are analyzed only after the analysis procedure has been finalized and validated with MC simulation.

%----------------------------------------------------
\section{BESIII detector and Monte Carlo simulation}
%----------------------------------------------------
The BESIII detector is located at the south collision point of the BEPCII~\cite{BEPC_II} storage ring, which operates at a center-of-mass energy range of $\sqrt{s}$ = 1.84 to 4.95~${\rm GeV}$, with a maximum luminosity of $1.1 \times 10^{33} ~\rm cm^{-2}s^{-1}$ at 3.773 ${\rm GeV}$.

The cylindrical core of the BESIII detector covers 93\% of the full solid angle and consists of a helium-based multilayer drift chamber (MDC), a plastic scintillator time-of-flight system (TOF), and a {CsI(Tl)} electromagnetic calorimeter (EMC). These components are enclosed within a superconducting solenoidal magnet that provides a 1.0 T magnetic field (0.9 T in 2012). The solenoid is supported by an octagonal flux-return yoke, with resistive plate counter muon identifier modules interleaved with steel. The BESIII coordinate system is defined such that the z-axis aligns with the beam direction, the x-axis is horizontal and points outward from the interaction point, and the y-axis is vertical, forming a right-handed coordinate system.

At a momentum of 1 ${\rm GeV}/c$, the charged particle momentum resolution is 0.5\%, and the specific ionization energy loss (dE/dx) resolution is 6\% for electrons from Bhabha scattering. The EMC has a photon energy resolution of 2.5\% (5\%) at 1 ${\rm GeV}$ in the barrel (end-cap) region. The time resolution in the TOF barrel region is 68 ps, while in the end-cap region it is 110 ps. In 2015, the end-cap TOF system was upgraded using multigap resistive plate chamber technology, improving the time resolution to 60 ps~\cite{etof}. Approximately 87\% of the data used in this analysis benefit from this upgrade.

Simulated samples produced with the {\sc geant4}-based~\cite{GEANT4}
Monte Carlo (MC) program which includes the geometric
description~\cite{detector_description1,detector_description2} of the
BESIII detector and the detector response, are used to determine the
detection efficiency and estimate the backgrounds. The simulation
includes the beam energy spread and initial state radiation in the
$e^+e^-$ annihilations modeled with the generator {\sc
  kkmc}~\cite{KKMC}. The inclusive MC sample includes both the
production of the $J/\psi$ resonance and the continuum processes
incorporated in {\sc kkmc}~\cite{KKMC}. All particle decays are
modelled with {\sc evtgen}~\cite{EVTGEN} using branching fractions
either taken from the Particle Data Group~\cite{PDG}, when available,
or otherwise estimated with {\sc
  lundcharm}~\cite{LUNDCHARM}. Final state radiation (FSR) from
charged final state particles is incorporated using the {\sc photos}
package~\cite{PHOTOS}. In this work, a sample of 100,000 phase space
(PHSP) MC events are generated to study the detection efficiencies for
the LNV decay, with equal contributions from each conjugate
channel. For background analysis, 10 billion inclusive MC events are used. The continuum background is estimated using off-resonance data samples at $\sqrt{s} = 3.650$, $3.682$ and $3.080$ ${\rm GeV}$~\cite{N3650}, with integrated luminosities of $410\ {\rm pb^{-1}}$, $404\ {\rm pb^{-1}}$ and $167\ {\rm pb^{-1}}$, respectively.

%----------------------------------------------------
\section{DATA ANALYSIS} \label{EVTSELET}
%----------------------------------------------------
%
% Track Level ==> Charged track pre-selection
%

All charged tracks are required to satisfy a geometrical acceptance of $|\rm{cos}\theta| < 0.93$, where $\theta$ is the polar angle of the charged track, defined with respect to the $z$-axis,
which is the symmetry axis of the MDC.  Each track must originate from the interaction region, defined as $R_{xy} <1.0$ cm and $|R_{z}| < 10.0$ cm, where $R_{xy}$ and $R_{z}$ are the distances of the closest approach to the interaction point of the track in the transverse plane and $z$-axis, respectively. Events with at least four charged tracks with zero net charge are retained for further analysis.

%
% Track Level ==> Charged track PID
%
Particle identification~(PID) for charged tracks combines measurements of the energy deposited in the MDC~(d$E$/d$x$), the flight time in the TOF, and the energy and shape of clusters in the EMC to calculate the confidence level (CL) for electron, pion, kaon hypotheses (CL$_e$, CL$_{\pi}$, CL$_K$).
Electron candidates must have CL$_e$ greater than 0.001, and the ratio of CL$_e$ to the sum of CLs for electrons, kaons, and pions (CL$_e$ + CL$_K$ + CL$_\pi$) should be greater than 0.8. In addition,  the ratio of energy deposited in the EMC($E$) to the momentum measured in the MDC($p$) must lie between 0.8$c$ and 1.2$c$.
For kaon candidates, the criteria include $\textrm{CL}_K>0$ and $\textrm{CL}_K>\textrm{CL}_{\pi}$. To further reduce the misidentification(mis-PID) between electrons and kaons, an additional requirement of $E/p$ less than 0.8$c$ is applied~\cite{ep_kaon}. The charges of the two selected kaons must be identical and opposite to the charges of the two selected electrons, which are also required to be the same. The minimum number of kaons and electrons are two. If there are multiple candidates in an event, the combination with the smallest $\chi^2_{4\rm C}$ is retained for further analysis, where the $\chi^2_{4\rm C}$ is from a kinematic fit~\cite{fit} enforcing four-momentum conservation (4C).

% Event Level ==> Kinematic Fit
To suppress potential contamination from mis-PID in processes with the same charged multiplicity, we perform an additional 4C kinematic fit with the hypotheses of  $e^+e^-\to$ $K^{+}K^{-}\pi^{+}\pi^{-}$, $K^{+}K^{-}K^{+}K^{-}$ and $\pi^{+}\pi^{-}\pi^{+}\pi^{-}$. By requiring that the $\chi^2_{4\rm C}$ of the $K^+K^+e^-e^-$ mass assignment is the smallest one among all possible assignments, such backgrounds can be effectively suppressed and only events with $\chi^2_{K^+K^+e^-e^-,4\rm C}<20$ are kept for further analysis. This requirement has been optimized with the Punzi significance~\cite{Punzi} using the formula $\frac{\varepsilon}{1.5+\sqrt{b}}$, where $\varepsilon$ is the detection efficiency from the signal MC simulation and $b$ is the number of background events from the inclusive MC sample. Furthermore, the selected $K^{+}K^{-}e^{+}e^{-}$ and $\pi^{+}\pi^{-}e^{+}e^{-}$ combinations will be rejected if the $e^+e^-$ pair comes from gamma conversions in the beam pipe.

% The invariant mass spectrum of $KKee$ ($M_{KKee}$) is fitted with a
% double Gaussian function for the signal and a second-order Chebyshev
% polynomial for the background.
The $J/\psi$ signal region is defined as $[3.07,~3.12]$ GeV/$c^2$, which corresponds to a range of three times of the mass resolution around the $J/\psi$ mass~\cite{PDG}. The detection efficiency is determined to be $15.68\%$ based on the simulated  $J/\psi \to K^+K^+e^-e^-$ events.

\begin{figure}[htbp]
\begin{center}
\includegraphics[scale=0.45]{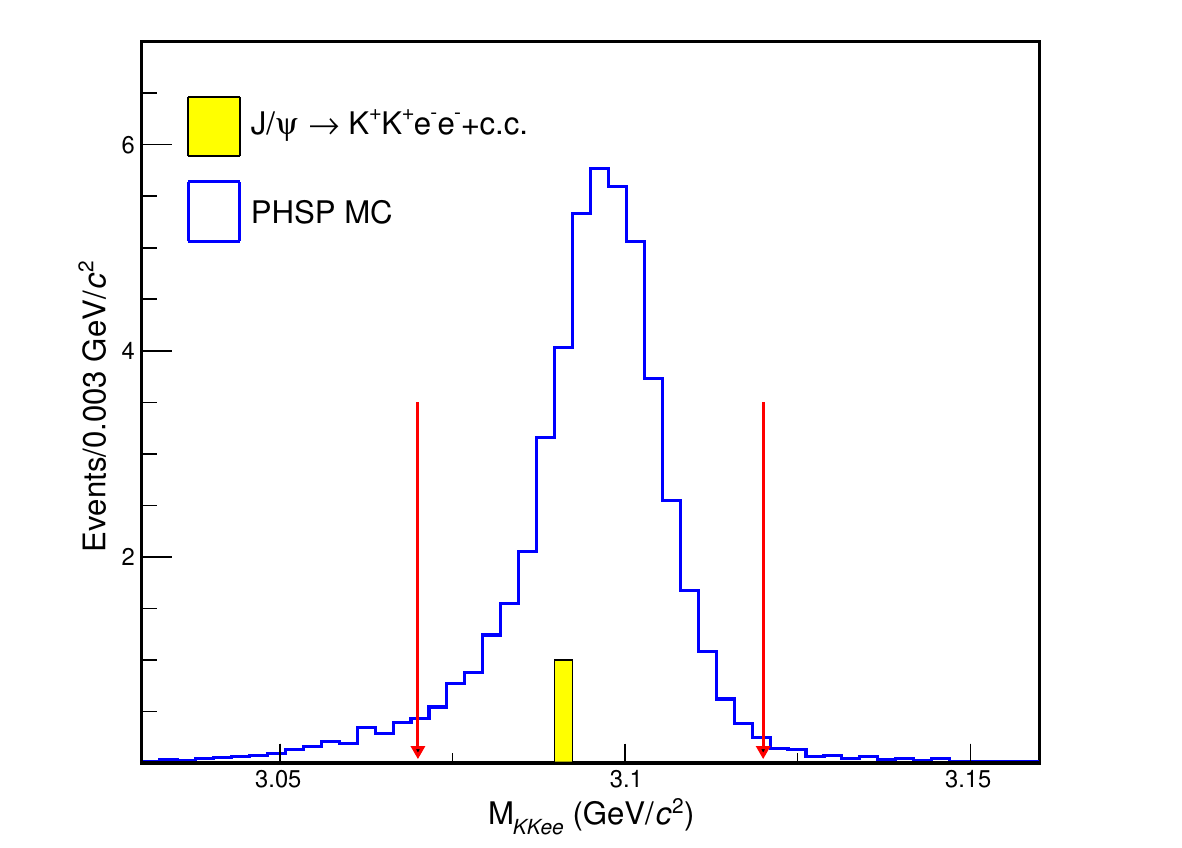}
    \caption{The distribution of $M_{K^{+}K^{+}e^{-}e^{-}+c.c.}$, where the blue histogram represents the simulated shape of signal, the yellow filled histogram represents data, and the red arrows indicate the signal region.}
    \vspace{0.05\baselineskip}
    \label{fig:data}
\end{center}
\end{figure}

%
% Background Contamination
%
With all above selections applied, only one event from data is left in the signal region as shown in Fig.~\ref{fig:data}; while no event survives from the inclusive MC sample. To investigate the potential contamination from continuum processes, we analyze the off-resonance data samples taken at $\sqrt{s}= 3.650$ GeV, $3.682$ GeV and $3.080$ GeV~\cite{N3650}. After considering the differences in integrated luminosities, cross sections, particle momenta, and center-of-mass energies, no event survives in the signal region, $N^{\rm continuum}=0$.  Thus, the total number of background events is calculated as $N^{\rm bkg}=N^{\rm inc}+N^{\rm continuum}=0$.

%
% Event Level ==> signal region && Detection Efficiency
%

%----------------------------------------------------
\section{Systematic Uncertainties}
%----------------------------------------------------
Systematic uncertainties in the measurement arise from the total number of $J/\psi$ events, tracking, PID, kinematic fit, MC modeling, and signal region.

\begin{itemize}
\item The uncertainty in the total number of $J/\psi$ events, determined using inclusive hadronic events, is 0.5\%~\cite{ref::jpsi_num}.
\item The uncertainty in the MDC tracking is estimated by comparing efficiency between data and MC  control samples of $J/\psi \to e^+ e^- (\gamma_{\mathrm{FSR}})$ and $J/\psi \to \pi^0 K^+ K^-$. The resulting uncertainties are 0.7\% per electron and 0.3\% per kaon. Consequently, the total systematic uncertainty in the MDC tracking is 2.0\%.
\item The uncertainty in the PID is evaluated using the same high-purity control samples of electrons and kaons as mentioned above. The uncertainties are found to be 0.5\% per electron and 0.6\% per kaon, resulting in a total PID uncertainty of 2.2\%.
\item The systematic uncertainty from the 4C kinematic fit and the $\chi^2_{4\rm C} < 20$ requirement is studied using a control sample of $J/\psi \to p \Bar{p} \pi^+ \pi^-$. The relative difference in efficiencies between data and MC simulation is found to be 6.2\%, which is taken as the systematic uncertainty from this source..
\item The systematic uncertainty associated with the signal region is also calculated using the control sample of $J/\psi \to p \Bar{p} \pi^+ \pi^-$. The relative difference in efficiencies, 0.4\%, is taken as the uncertainty.
%
% The MC statistics uncertainty is from the calculation of signal efficiency. It is the limitation of the signal MC sample size that contributes to the MC statistics uncertainty. The error of detection efficiency is 0.9\%.
 \item The uncertainty due to the signal MC sample size is 0.9\%.
%----------------------------------------------------------------
\item To obtain the systematic uncertainty from the MC modeling, we compare the difference of the detection efficiency between the PHSP generator and two alternative MC models, i.e., $J/\psi\to\nu_M K^+e^-\to (K^+e^-)K^+e^-$ and $J/\psi\to\nu_M\nu_M\to(K^+e^-)(K^+e^-)$, where the Majorana neutrino $\nu_M$ decays into $K^+e^-$ with different Majorana masses and corresponding detection efficiencies.
The Majorana neutrino mass is varied from the $Ke$ mass threshold to the maximum phase space allowed by the $J/\psi$ decay, approximately  from 0.575 GeV/$c^2$ to 1.525 GeV/$c^2$, with an interval of 0.025 GeV/$c^2$. The largest efficiency difference, 1.0\%, is taken as the systematic uncertainty.

\end{itemize}

All the uncertainties are summarized in Table~\ref{tab::sys}, and the total systematic uncertainty~($\Delta^{\rm sys}$) is obtained by adding the individual effects in quadrature.

\begin{table}[h!]
\centering
\caption{Summary of the systematic uncertainties.}
\begin{tabular}{lc}
\hline
\hline
Source & Uncertainty~(\%)\\
\hline
$N_{J/\psi}$ 		& 0.5\\
MDC tracking 		& 2.0\\
PID 				& 2.2\\
4C kinematic fit and $\chi^2$ cut 	& 6.2\\
signal region 		& 0.4\\
MC statistics 		& 0.9\\
MC modeling 	& 1.0\\
\hline
Total $\Delta^{\rm sys}$	& 7.0		\\
\hline
\hline
\end{tabular}
\label{tab::sys}
\end{table}

%----------------------------------------------------
\section{Results}
%----------------------------------------------------

\subsection{Upper limit on  $J/\psi \to K^+K^+e^-e^-$}
Since one event is observed in the signal region, the signal yield ($N^{\rm sig}$) is set to one and the background yield ($N^{\rm bkg}$) is set to be zero. The efficiency-corrected upper limit (UL) on the signal yield $N^{\rm up}_{K^+K^+e^-e^-}$ is estimated to be $23.2$ at the 90\% CL using a frequentist method \cite{TROLKE}. This estimation employs an unbounded profile
likelihood approach to account for systematic uncertainties
($\Delta^{sys}=7.0\%$), where both signal events ($N^{sig}=1$) and
background events ($N^{bkg}=0$) modeled by a Poisson distribution,  the detection efficiency ($\varepsilon_{K^+K^+e^-e^-}=15.83\%$) following a Gaussian distribution, with the systematic uncertainty as its standard deviation.  The UL of the branching fraction (BF)  of ${J/\psi}\to K^{+}K^{+}e^{-}e^{-}$ at the 90\% CL is determined by
\begin{linenomath*}
\begin{equation}
\begin{aligned}
    \mathcal{B}(J/\psi \to K^+K^+e^-e^-)
    &< \mathcal{B}^{\rm UL}(J/\psi \to K^+K^+e^-e^-) \\
    &= \frac{N^{\mathrm{up}}_{K^+K^+e^-e^-}}{N^{\mathrm{tot}}_{J/\psi}} \\
    &= 2.1 \times 10^{-9},
    \label{BF}
\end{aligned}
\end{equation}
\end{linenomath*}
%\begin{linenomath*}
%\begin{align}
%	\mathcal{B}(J/\psi \to K^+K^+e^-e^-)
%	<& \mathcal{B}^{\rm UL}(J/\psi \to K^+K^+e^-e^-)  \notag \\
%	&=\frac{N^{\mathrm{up}}_{K^+K^+e^-e^-}}{N^{\mathrm{tot}}_{J/\psi}} \notag \\
%	&= 2.1\times10^{-9},
 %   \label{BF}
%\end{align}
%\end{linenomath*}
where $N_{J/\psi}^{{\rm tot}} = (10087\pm 44)\times10^{6}$ is the total number of $J/\psi$ events in data.

\subsection{Majorana mass dependent upper limits}
In addition to the overall UL, assuming the LNV signal from Majorana neutrino, we also investigate the ULs on the branching fractions of $J/\psi \to 2\nu_M\to (K^+e^-)(K^+e^-)$ with different Majorana masses. Full kinematic ranges of the $Ke$ invariant mass ($m_{Ke}$) from the threshold to the largest phase space allowed in the $J/\psi$ decay,
\begin{linenomath*}
\begin{equation}
m_K+m_e \leq m_{\nu_M} \leq \frac{m_{J/\psi}}{2},
\label{ke mass}
\end{equation}
\end{linenomath*}
is considered. We take the Majorana neutrino mass $m_{\nu_M}$ from 0.575 GeV/$c^2$ to 1.500 GeV/$c^2$ and divide this mass range into uniform intervals, with a width of 0.025 GeV/$c^2$. In each interval, the Majorana mass is taken as the center value.
Due to extremely small phase space, the first three (0.500 GeV/$c^2$, 0.525GeV/$c^2$, and 0.550 GeV/$c^2$) are not selected as mass possibilities.

\begin{figure}[ht]
\centering
\includegraphics[scale=0.45]{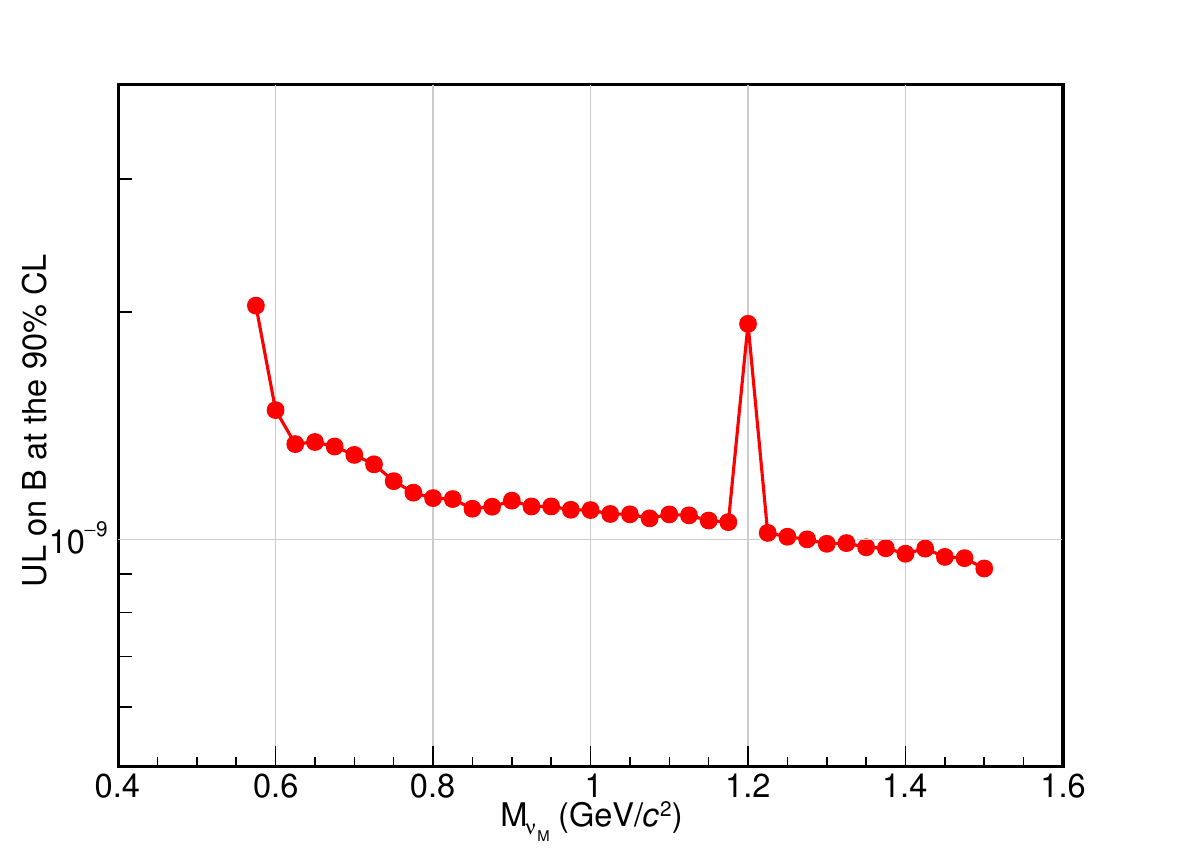}
\caption{The ULs at the 90\% CL for the data sample as a function of $m_{\nu_M}$ for the decay $J/\psi \to 2\nu_M\ (\to K^+e^-)$.}
%\vspace{0.05\baselineskip}
\label{fig::ULsSemi}
\end{figure}

The values of the mass dependent detection efficiencies, the signal event
counts, the background event counts, the total systematic uncertainties,
and the ULs on the BFs at the 90\% CL are reported in
Table~\ref{tab::result}, for each $m_{Ke}$ bin. The mass dependent ULs
are calculated to be from [2.04, 0.92]$\times10^{-9}$ with the same
method, and the results are shown in Fig.~\ref{fig::ULsSemi}.

\begin{table}
\small
% \footnotesize

	\centering
	\caption{The mass dependent detection efficiencies($\epsilon_{i}$), the signal event counts($N^{\rm sig}_{i}$), the background event counts($N^{\rm bkg}_{i}$), the total systematic uncertainties ($\Delta^\mathrm{sys}_\mathrm{i}$), and the ULs on the BFs at the 90\% CL($\mathcal{B}^\mathrm{UL}$) for different Majorana mass hypotheses from 0.575 GeV/$c^2$ to 1.525 GeV/$c^2$ with an interval of 0.025 GeV/$c^2$.}
	\begin{tabular}{c|rrrrr}
		\hline
		\hline
		$m_{\nu_M}$(GeV/$c^2$) &\multicolumn{1}{c}{$\varepsilon_{2\nu_M}$~(\%)} &\multicolumn{1}{c}{$N^{\rm sig}_{i}$} &\multicolumn{1}{c}{$N^{\rm bkg}_{i}$} &$\Delta^{\rm sys}_{i}$~(\%) &\multicolumn{1}{c}{$\mathcal{B}^\mathrm{UL}$~($10^{-9}$)}  \\

		\hline
            0.575 & 8.97 	&\multicolumn{1}{c}{0} &\multicolumn{1}{c}{0}  &26.8 		&$2.04$	\\
		0.600 &12.34 	&\multicolumn{1}{c}{0} &\multicolumn{1}{c}{0}  &25.8 		&$1.48$	\\
		0.625 &13.69 	&\multicolumn{1}{c}{0} &\multicolumn{1}{c}{0}  &26.4 		&$1.34$	\\
		0.650 &13.60 	&\multicolumn{1}{c}{0} &\multicolumn{1}{c}{0} & 22.9  	&$1.35$	\\
		0.675 &13.79 	&\multicolumn{1}{c}{0} &\multicolumn{1}{c}{0}  &19.3 		&$1.33$	\\
		0.700 &14.15 	&\multicolumn{1}{c}{0} &\multicolumn{1}{c}{0} & 16.9 	    &$1.29$	\\
		0.725 &14.56 	&\multicolumn{1}{c}{0} &\multicolumn{1}{c}{0} & 15.5  	&$1.26$	\\
		0.750 &15.32 	&\multicolumn{1}{c}{0} &\multicolumn{1}{c}{0} & 14.1  	&$1.19$	\\
		0.775 &15.87 	&\multicolumn{1}{c}{0} &\multicolumn{1}{c}{0} & 12.1 		&$1.15$	\\
		0.800 &16.14 	&\multicolumn{1}{c}{0} &\multicolumn{1}{c}{0} & 11.3 		&$1.13$	\\
		0.825 &16.18 	&\multicolumn{1}{c}{0} &\multicolumn{1}{c}{0} & 8.4 		&$1.13$	\\
		0.850 &16.67 	&\multicolumn{1}{c}{0} &\multicolumn{1}{c}{0} & 11.2 		&$1.10$	\\
		0.875 &16.57 	&\multicolumn{1}{c}{0} &\multicolumn{1}{c}{0} & 9.0 		&$1.11$	\\
		0.900 &16.26 	&\multicolumn{1}{c}{0} &\multicolumn{1}{c}{0} & 7.0 		&$1.11$	\\
		0.925 &16.55 	&\multicolumn{1}{c}{0} &\multicolumn{1}{c}{0} & 7.0 		&$1.11$	\\
		0.950 &16.55 	&\multicolumn{1}{c}{0} &\multicolumn{1}{c}{0} & 7.0 		&$1.09$	\\
		0.975 &16.72 	&\multicolumn{1}{c}{0} &\multicolumn{1}{c}{0}  & 7.0 		&$1.09$	\\
		1.000 &16.74 	&\multicolumn{1}{c}{0} &\multicolumn{1}{c}{0} & 7.0 		&$1.08$	\\
		1.025 &16.94 	&\multicolumn{1}{c}{0} &\multicolumn{1}{c}{0} & 7.1 		&$1.08$	\\
		1.050 &16.96 	&\multicolumn{1}{c}{0} &\multicolumn{1}{c}{0} & 7.0 		&$1.07$	\\
		1.075 &17.16 	&\multicolumn{1}{c}{0} &\multicolumn{1}{c}{0} & 7.2 		&$1.08$	\\
		1.100 &16.96 	&\multicolumn{1}{c}{0} &\multicolumn{1}{c}{0} & 7.5 		&$1.08$	\\
		1.125 &17.01 	&\multicolumn{1}{c}{0} &\multicolumn{1}{c}{0} & 7.8 		&$1.06$	\\
		1.150 &17.28 	&\multicolumn{1}{c}{0} &\multicolumn{1}{c}{0} & 7.0 		&$1.05$	\\
		1.175 &17.36 	&\multicolumn{1}{c}{0} &\multicolumn{1}{c}{0} & 7.4 		&$1.05$	\\
		1.200 &17.47 	&\multicolumn{1}{c}{1} &\multicolumn{1}{c}{0} & 7.2 		&$1.93$	\\
		1.225 &17.94 	&\multicolumn{1}{c}{0} &\multicolumn{1}{c}{0} & 7.0 		&$1.02$	\\
		1.250 &17.16 	&\multicolumn{1}{c}{0} &\multicolumn{1}{c}{0} & 7.0 		&$1.01$	\\
		1.275 &18.30 	&\multicolumn{1}{c}{0} &\multicolumn{1}{c}{0} & 7.5 		&$1.00$	\\
		1.300 &18.55 	&\multicolumn{1}{c}{0} &\multicolumn{1}{c}{0} & 7.3 		&$0.99$	\\
		1.325 &18.51 	&\multicolumn{1}{c}{0} &\multicolumn{1}{c}{0} & 7.6 		&$0.99$	\\
		1.350 &18.75 	&\multicolumn{1}{c}{0} &\multicolumn{1}{c}{0} & 8.5 		&$0.98$	\\
		1.375 &18.80 	&\multicolumn{1}{c}{0} &\multicolumn{1}{c}{0} & 8.9 		&$0.97$	\\
		1.400 &19.12 	&\multicolumn{1}{c}{0} &\multicolumn{1}{c}{0} & 9.4 		&$0.96$	\\
		1.425 &18.82 	&\multicolumn{1}{c}{0} &\multicolumn{1}{c}{0} &8.7 		&$0.97$	\\
		1.450 &19.31 	&\multicolumn{1}{c}{0} &\multicolumn{1}{c}{0} &11.5 		&$0.95$	\\
		1.475 &19.38 	&\multicolumn{1}{c}{0} &\multicolumn{1}{c}{0} &11.1 		&$0.95$	\\
		1.500 &20.00 	&\multicolumn{1}{c}{0} &\multicolumn{1}{c}{0} &12.8 		&$0.92$	\\
                1.525 &20.28 	&\multicolumn{1}{c}{0} &\multicolumn{1}{c}{0} &16.2 		&$0.90$	\\
		\hline
		\hline
	\end{tabular}
	\label{tab::result}
\end{table}

%----------------------------------------------------
\section{Summary}
%----------------------------------------------------
In summary, by using $(10087\pm 44)\times10^{6}$ $J/\psi$ events collected by the BESIII detector, we search for the LNV decay $J/\psi\to K^+K^+e^-e^- +c.c.$ for the first time. One event survives in the signal region and the UL on the decay branching fraction of $J/\psi\to K^+K^+e^-e^- +c.c.$ is determined to be $2.1 \times 10^{-9}$ at the 90\% CL. The Majorana mass dependent ULs on the branching fractions at the 90\% CL are estimated, varying within the range of [0.92, 2.04]$\times10^{-9}$. By utilizing these results, constraints on the mixed matrix element $|V_{e\nu_{M}}|^2$ between the Majorana neutrino and electron can be derived within the theoretical models.

%----------------------------------------------------
\section{Acknowledgements}
%----------------------------------------------------
The BESIII Collaboration thanks the staff of BEPCII (https://cstr.cn/31109.02.BEPC) and the IHEP computing center for their strong support. This work is supported in part by National Key R\&D Program of China under Contracts Nos. 2023YFA1606000, 2023YFA1606704; National Natural Science Foundation of China (NSFC) under Contracts Nos. 12035009, 11635010, 11935015, 11935016, 11935018, 12025502, 12035013, 12061131003, 12192260, 12192261, 12192262, 12192263, 12192264, 12192265, 12221005, 12225509, 12235017, 12361141819; the Chinese Academy of Sciences (CAS) Large-Scale Scientific Facility Program; CAS under Contract No. YSBR-101; 100 Talents Program of CAS; The Institute of Nuclear and Particle Physics (INPAC) and Shanghai Key Laboratory for Particle Physics and Cosmology; German Research Foundation DFG under Contract No. FOR5327; Istituto Nazionale di Fisica Nucleare, Italy; Knut and Alice Wallenberg Foundation under Contracts Nos. 2021.0174, 2021.0299; Ministry of Development of Turkey under Contract No. DPT2006K-120470; National Research Foundation of Korea under Contract No. NRF-2022R1A2C1092335; National Science and Technology fund of Mongolia; National Science Research and Innovation Fund (NSRF) via the Program Management Unit for Human Resources \& Institutional Development, Research and Innovation of Thailand under Contract No. B50G670107; Polish National Science Centre under Contract No. 2024/53/B/ST2/00975; Swedish Research Council under Contract No. 2019.04595; U. S. Department of Energy under Contract No. DE-FG02-05ER41374.

\end{document}